\batchmode
\documentclass[conference,onecolumn]{IEEEtran}
\usepackage[T1]{fontenc}
\usepackage[latin9]{inputenc}
\usepackage{amsmath}
\usepackage{amsthm}
\usepackage{amssymb}
\usepackage{graphicx}
\usepackage[pdftex,unicode=true,
 bookmarks=true,bookmarksnumbered=true,bookmarksopen=true,bookmarksopenlevel=1,
 breaklinks=false,pdfborder={0 0 0},pdfborderstyle={},backref=false,colorlinks=false]
 {hyperref}
\hypersetup{pdftitle={Your Title},
 pdfauthor={Your Name},
 pdfpagelayout=OneColumn, pdfnewwindow=true, pdfstartview=XYZ, plainpages=false}

\makeatletter
\theoremstyle{definition}

\theoremstyle{plain}

\ifx\proof\undefined\
  \newenvironment{proof}[1][\proofname]{\par
    \normalfont\topsep6\p@\@plus6\p@\relax
    \trivlist
    \itemindent\parindent
    \item[\hskip\labelsep
          \scshape
      #1]\ignorespaces
  }{%
    \endtrivlist\@endpefalse
  }
  \providecommand{\proofname}{Proof}
\fi
\theoremstyle{remark}

\usepackage[caption=false,font=footnotesize]{subfig}
\usepackage{mathtools}
\usepackage{scrlayer-scrpage}
\clearpairofpagestyles
\@ifundefined{showcaptionsetup}{}{%
 \PassOptionsToPackage{caption=false}{subfig}}
\usepackage{subfig}
\makeatother

\providecommand{\examplename}{Example}
\providecommand{\propositionname}{Proposition}
\providecommand{\remarkname}{Remark}

\begin{document}
\title{On-Demand Video Dispatch Networks:\\
A Scalable End-to-End Learning Approach\vspace{-5mm}
}
\author{\IEEEauthorblockN{Damao Yang}
\IEEEauthorblockA{Bilibili Inc. \\ yangdamao@bilibili.com \\ adrian.y.dm@gmail.com}
\and
\IEEEauthorblockN{Sihan Peng}
\IEEEauthorblockA{NYU Shanghai \\ sp4059@nyu.edu}
\and
\IEEEauthorblockN{He Huang}
\IEEEauthorblockA{Purdue University \\ hhuang1206@gmail.com}
\and
\IEEEauthorblockN{Hongliang Xue}
\IEEEauthorblockA{Cornell University \\ xuehongliang518@gmail.com}
}
\IEEEspecialpapernotice{}
\IEEEaftertitletext{}
\maketitle
\begin{abstract}

We design a dispatch system to improve the peak service quality of video on demand (\textbf{VOD}). Our system predicts the hot videos during the peak hours of the next day based on the historical requests, and dispatches to the content delivery networks (\textbf{CDN}s) at the previous off-peak time. In order to scale to billions of videos, we build the system with two neural networks, one for video clustering and the other for dispatch policy developing. The clustering network employs autoencoder layers and reduces the video number to a fixed value. The policy network employs fully connected layers and ranks the clustered videos with dispatch probabilities. The two networks are coupled with weight-sharing temporal layers, which analyze the video request sequences with convolutional and recurrent modules. Therefore, the clustering and dispatch tasks are trained in an end-to-end mechanism. The real-world results show that our approach achieves an average prediction accuracy of 17\%, compared with 3\% from the present baseline method, for the same amount of dispatches.

\end{abstract}

\section{Introduction and Related Work}

A typical architecture of Internet VOD is depicted in Figure 1(a). Given the knowledge of the relation between users and videos as well as of the relation between users and CDNs, one key issue as Figure 1(b) is to dispatch videos to CDNs. Improper dispatches cause poor video service qualities. It is noticed that five main challenges occur with the emerging popularity of VOD: 
(1) large video amount (currently 50 million) compared with limited storage capacity (800,000 videos per CDN on average) and dispatch bandwidth (30,000 videos per day on average); 
(2) ever-increasing uploaded videos (60,000 videos per day on average) every day; 
(3) notable differences between peak (1,667,000 qps) and off-peak (235,000 qps) requests and video preferences; 
(4) erratic variants of user requests; 
(5) implicit and intrinsic request relations among users due to different populations, economies and cultures.  
Bearing the above challenges in mind, we aim at a scalable video dispatch system working at off-peak time to enhance the peak service qualities on the next day.  

\begin{figure}[ht]
  \centering
  \includegraphics[width=0.5\textwidth]{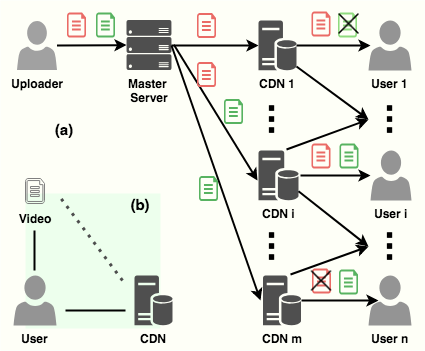}
  \caption{Video Dispatch System and Problems}
\end{figure}

Our present baseline method triggered the dispatch of video $v$ to CDN $i$, when the requests of $v$ from the users served by $i$ exceed a threshold $h$ for a period $p$. This implies that one successful service corresponds to h CDN misses. Besides, dispatch at peak time occupies network bandwidth and resources, which affects the peak service quality. We apply two ratios as evaluation metric: $r_{whole}$ as the number of requested videos after dispatched within a whole day over the total dispatch number, and $r_{peak}$ as the number of peak requested dispatched from several hours ago till recently over the total dispatch number. While $r_{whole}$ can be optimized by fine tuning $h$ and $p$, $r_{peak}$ stays low, meaning that the baseline method performs poor for peak service quality. This is due to difficulty (3) and (4), and the baseline method fails to explore the relation between peak and off-peak requests, only to exploit a very short memory of requests without knowing the longer changing pattern.  

\subsection{Time Series Forecasting}

Recurrent neural networks (\textbf{RNN}s) memorize state transition efficiently thanks to their recursive structures with hidden state storage. RNN has been applied to predict noisy and non-stationary time series, where the series are unsupervised learned before sent to RNN in case that RNN prefers short-term dependencies than long-term dependencies caused by vanishing gradients. Long short-term memories (\textbf{LSTM}s) and gated recurrent units (\textbf{GRU}s) are proposed with learnable gates to control information flowing in order to alleviate this problem. Both LSTM and GRU have achieved the state-of-the-art performances in sequential applications. A common point in the above sequential models is that every single point in the series weighs equally importantly, and therefore they are vulnerable to noise and prone to overfitting to specific series patterns.

Convolutional neural networks (\textbf{CNN}s), which are typically implemented with globally shared nonlinear filters followed by pooling operations, strengthen and extract local features. CNN has formed the crux of computer vision applications. CNN is usually followed with max pool to subtract the most representative features for each local region. 

Time series analysis with CNN (Borovykh et al. 2017) also emerges in recent years. In this paper, we put a top-k pool ahead of CNN to extract the peak-time pattern, place three max pools behind CNN to extract the inter-day pattern, and concatenate CNN with RNN (Jozefowicz et al. 2016) to learn the dynamics with low complexity and in avoidance of overfitting compared with pure RNN. In Wavenet, causal CNN without pools (reference ?) were proposed to generate predicted 1D signals, forcing the filters always to calculate on the previous points.

It is noticeable that in the above researches, all the input and output elements value equally. However, the peak-time-oriented dispatch problem focuses more on the peak elements than the off-peak ones. Besides, the predicted peak requests are not necessarily precise in an exact time, since the peak time sustains in a period. This motivates us to design our algorithm to differentiate peak and off-peak request patterns.

\subsection{Ranking Problems}
At first glance, video dispatch can be solved the same way as a variation of 0/1 knapsack problem. Metaheuristic techniques are guaranteed neither an optimal solution, nor polynomial prediction time complexity. Neural network methods, such as Hopfield Net \cite{Hopfield1985}, and recently proposed Pointer Net \cite{Vinyals2015}, though have been successfully verified in similar combinatorial optimization applications, suffer from either scalability or low-dimension restrictions.

Alternatively, the dispatch problem is to pick top-k videos by the probabilities of the future requests from the users served by certain CDN servers. This process simulates a ranking problem. Learning to rank (\textbf{L2R}) has produced a much better performance than traditional methods. The L2R algorithms are categorized \cite{liu2009} as pointwise \cite{li2008mcrank}, and other techniques. The pointwise ranking focuses on total and absolute orders. This works for the dispatch problem because the strategies cannot be made until the total video requests have been seen, while the pairwise ranking aims at partial orders and the listwise ranking costs too high complexity for large scale data set.

\subsection{Clustering and Manifold Learning}

In the ranking or time series forecasting problems, the set of ranked items or series are predetermined. The VOD provider, however, is facing new videos all the time, which is nontrivial in dispatch problems. Clustering methods categorize data with distance-based metric without human labelling so that new videos as well as present ones are represented with a nearest centroid. To cope with ``curse of dimensionalit'', manifold learning projects features into a low dimensional subspace, uncovering the intrinsic distribution of the input data in visible and ``compact'' way. Autoencoder \cite{hinton2006}, as a nonlinear dimensionality reduction method, was proposed to pretrain deep neural networks. It and variations \cite{kingma2013,MakhzaniSJG15} have been applied as preprocessing for clustering \cite{baldi2012,xie2016}.

Distributed clustering algorithms \cite{tasoulis2004} were proposed for large data set. Unfortunately, they do not maintain consistent clustering indices at training and predicting iterations. We propose a supervised-like clustering modification to solve this problem.

\subsection{End-to-end Learning}

Recently it has been a fashion to train multi-task neural networks in an end-to-end learning mechanism for better performances rather than to optimize them independently. Within the ideal end-to-end learning trait, all learnable parameters are differentiable with respect to the ultimate task`s losses (in supervised learning) or rewards (in reinforcement learning). One strategy is to carefully design the structure of these networks so that they are connected by shared components. For example, Faster Rcnn \cite{ren2015} combined region proposal with objection recognition by a shared CNN part as feature extractors, and led to globally optimized object detection results; value iteration networks \cite{tamar2016} introduced a fully differentiable CNN planning module to approximate the value iteration algorithm, simultaneously approximating the Markov decision process and optimizing the control.

\subsection{Our Contributions}
\begin{enumerate}
\item An end-to-end learning mechanism which fulfills video clustering and dispatch tasks.
\item A novel structure for time series forecasting which utilizes both peak-time and whole-day temporal features.
\item A supervised-like clustering method fit for the mini-batch gradient descent (\textbf{MBGD}) training and the distributed environment, and therefore scalable for big data. 
\item State-of-the-art performances under the real-world VOD system.
\end{enumerate}

 \subsection{Notations}
The symbols in this paper are defined as follows:

$D$ denotes the constant as the number of consecutive days when training or predicting a batch of data.

$T$ denotes the constant as the number of time intervals within a day.

$K$ denotes the constant as the number of peak time intervals within a day.

$U$ denotes the set of the users.

$I$ denotes the set of CDNs, and $I_{u}$ the set of the CDNs that serve user $u$. 

$V$ denotes the set of the videos, and $V_i$ the set of the videos dispatched to CDN $i$

$R$ denotes the set of the requested videos within a day, $R_u$ the set of the requested videos from the user $u$ within a day, $R^d$ the set of the requested videos on day $d$, and $R^u_d$ the set of the requested videos by the user u on day $d$. $PR$ denotes the set of the requested videos at peak time, and similarly with $PR_u$, $PR_d$, and $PR^u_d$.

$C$ denotes the mapping from the video to its cluster, and $C_{(r)}$ denotes the cluster of the video $r$. $CP$ denotes the probabilities of the clustered videos dispatched to the CDNs, and $CP_{C(r),i}$ the probability of the cluster of the requested video $r$ dispatched to IDC $i$. $CPT$ denotes the upper bound of the sum of clustered dispatch probabilities to the CDNs.

$|A|$ denotes the cardinality of the set $A$.

$\Re(f)$ denotes the cardinality of the range of the mapping $f$.

$A1\times{A2}$ denotes the Cartesian product of two sets $A1$ and $A2$.

$\otimes$ represents convolution operator.

$top(K|S)$ represents extracting the $K$ largest elements out of $S$ according to the ordered set given by $S$. 

$sort(V)$ represents sorting $V$ in descending order.

$maxpool(M,r,c)$ represents to extract the max element of each submatrix $M_{r\times c}^{'}$ of the matrix $M$, and reshape the output to 1D.

$ReLU(x)$, $sigmoid(x)$, and $tanh(x)$ are the nonlinear activation functions, referring to rectifier, sigmoid, and hyperbolic tangent functions.

$||v||_2$ denotes the L2-norm of $v$.

\section{Problem Formulation}

\setlength\arraycolsep{2pt}
$$\min_{C, CP} \sum\limits_{u\in U} \sum\limits_{r\in PR_{u}^{D+1}} |r| \prod\limits_{i\in I_{u}} \left. \left( 1-CP_{C(r),i} \right) \right\vert \begin{matrix} R^1 & ... & R^D \end{matrix} $$
$$\emph{s. t.} \sum\limits_{r\in{\bigcup\limits_{u\in U} \bigcup\limits_{{d=1}}^D R_{u}^{d}}} CP_{C(r),i} \leq CPT_{i} , \forall i \in I$$
$$ 0.0 \leq CP_{C(r),i} \leq 1.0, \forall r \in \bigcup\limits_{u\in U} \bigcup\limits_{{d=1}}^D R_{u}^{d}, \forall i \in I$$
$$ C(r) \in C, \forall r \in \bigcup\limits_{u\in U} \bigcup\limits_{{d=1}}^D R_{u}^{d}, \forall i \in I$$

\section{The Structure of Dispatch Neural Networks}

We set up two neural networks, clustering network and policy network, as depicted in Figure 2. The clustering network is composed of temporal layers and clustering layers. The policy network comprises accumulation layers, temporal layers and policy layers. The two networks are coupled with the temporal layers, which have identical structures and shared weights. The outputs of the clustering network and the primitive inputs constitute the inputs of the policy network. The outputs of the policy network are the probabilities of the videos dispatched to users rather than to CDNs. The probabilities of the videos dispatched to CDNs will be calculated in a following ``post-processing'' procedure.

\begin{figure}[ht]
  \centering
  \includegraphics[width=0.4\textwidth]{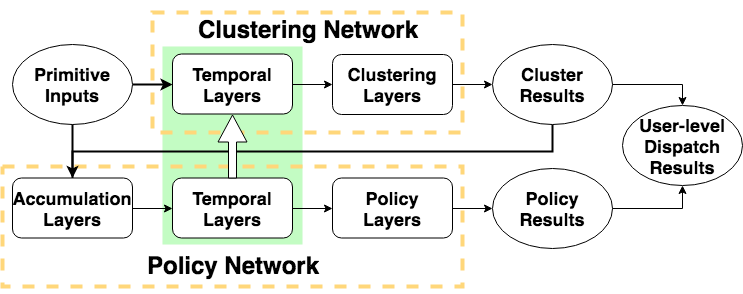}
  \caption{Structure of the Two Coupled Networks}
\end{figure}

\subsection{Primitive Inputs}
The primitive inputs for one video are its request sequences. Each sequence represents the requests from a user on $T$ intervals of $D$ days. Therefore, the primitive inputs have a user dimension and a time dimension with intraday and interday parts. The inputs of video $v$ is denoted as follows, where $x_u^{t,d} \geq 0$ denotes the requested number of $v$ from the user $u$ on interval $t$ on day $d$.  

$$X_{(v)} = \begin{pmatrix}
      \begin{pmatrix} x_1^{1,1} & ... & x_1^{T,1} \end{pmatrix} & ... & \begin{pmatrix} ( x_1^{1,D} & ... &  x_1^{T,D} \end{pmatrix} \\ 
      \vdots \\ 
      \begin{pmatrix} x_{|U|}^{1,1} & ... & x_{|U|}^{T,1} \end{pmatrix} & ... & \begin{pmatrix} ( x_{|U|}^{1,D} & ... &  x_{|U|}^{T,D} \end{pmatrix} 
   \end{pmatrix} _{(v)} $$

We define $X_u^{t,d} = \begin{pmatrix} x_u^{1,d} & ... & x_u^{T,d} \end{pmatrix} $ for convenience.

The primitive inputs exclude the information other than video requests, such as video metadata, weekends and holidays. There are three reasons: (1) the mapping from video content to user request is not bijection due to redundant uploads; (2) labelling the videos with tags may be subjective and ambiguous; (3) the video requests are highly nonstationary processes, and modeling with other observations is vulnerable to overfitting.

\subsection{Temporal Layers}
The temporal layers process the time dimension of the inputs. There are two components, CNN module for the intraday part and RNN module for the interday part, as depicted in Figure 3. The temporal layers are designed to predict the future requests with daily peak-time features and whole-day trajectories.

The CNN module has two parallel groups of convolutional filters with pools, which we call peak convolutional part (\textbf{PCP}) and mean convolutional part (\textbf{MCP}) respectively. PCP and MCP take $X_u^d$ as a batch. 

In PCP, the inputs are processed with a global top-K pool sorted by the total requests at each time. Then 1D convolutions are calculated. 
$$ XK_u^{d} = top \left( K \bigg| argsort \left( \sum\limits_{x \in V}  X_u^d \right) \right) $$

$$ YP_u^d = \emph{ReLU} (W_{peak} \otimes XK_u^d + b_{peak}) $$

$W_{peak}$ and $b_{peak}$ are learnable parameters. 

In MCP, the inputs are passed via 1D convolution filters.

$$ YMC_u^d = \emph{ReLU} (W_{mean} \otimes X_u^d + b_{mean}) $$ 

$W_{mean}$ and $b_{mean}$ are learnable parameters. 

Then $YMC_u^d$ is reshaped to three matrices $Row_u^d$, $Column_u^d$, $Block_u^d$: $R_{row} \times R_{column}$, $C_{row} \times C_{column}$, and $(B_{row1} \times B_{row2}) \times (B_{column1} \times B_{column2}) $, where $T = R_{row} \times R_{column} = C_{row} \times C_{column} = (B_{row1} \times B_{row2}) \times (B_{column1} \times B_{column2})$, and $K = R_{row} = C_{column} = (B_{row1} \times B_{column1})$
The rows are closely related to hours, and the columns to minutes.

These matrices are processed with 2D max pools and weightedly summed:
$$ YR_u^d = \emph{maxpool} (Row_u^d, 1, R_{column})$$
$$ YC_u^d = \emph{maxpool} (Column_u^d , C_{row}, 1)$$
$$ YB_u^d = \emph{maxpool} (Block_u^d, B_{row2}, B_{column2})$$
$$ YM_u^d = w_{row}  YR_u^d + w_{column}  YC_u^d + w_{block}  YB_u^d$$

$w_{row}, w_{column}$, and $w_{block}$ are learnable parameters.

The outputs of the CNN module, $C_u^d$, are the concatenation of $YP_u^d$ and $YM_u^d$.

The RNN module adopts GRU, and takes $C_u^d$ as the $d$th input for user $u$. The $d$th output $R_u^d$ is computed as :

$$ z_u^d = \emph{sigmoid} (W_z C_u^d + U_z R_u^{d-1} + b_z) $$
$$ r_u^d = \emph{sigmoid} (W_r C_u^d + U_r R_u^{d-1} + b_r) $$
$$ R_u^d = (1 - z_u^d)  R_u^{d-1} +
 z_u^d  \emph{tanh} (W_h C_u^d + U_h(r_u^d  R_u^{d-1}) + b_h) $$

$W_z$, $U_z$, $b_z$, $W_r$, $U_r$, $b_r$, $W_h$, $U_h$, and $b_h$ are learnable parameters.

The outputs of the temporal layers from user $u$ are $R_u^D$. We denote the outputs for video $v$ from all users as $T_{(v)} = \begin{pmatrix} R_1^D & ... & R_{|U|}^D \end{pmatrix} _{(v)}$. 

\begin{figure}[h]
  \centering
  \includegraphics[width=0.7\textwidth]{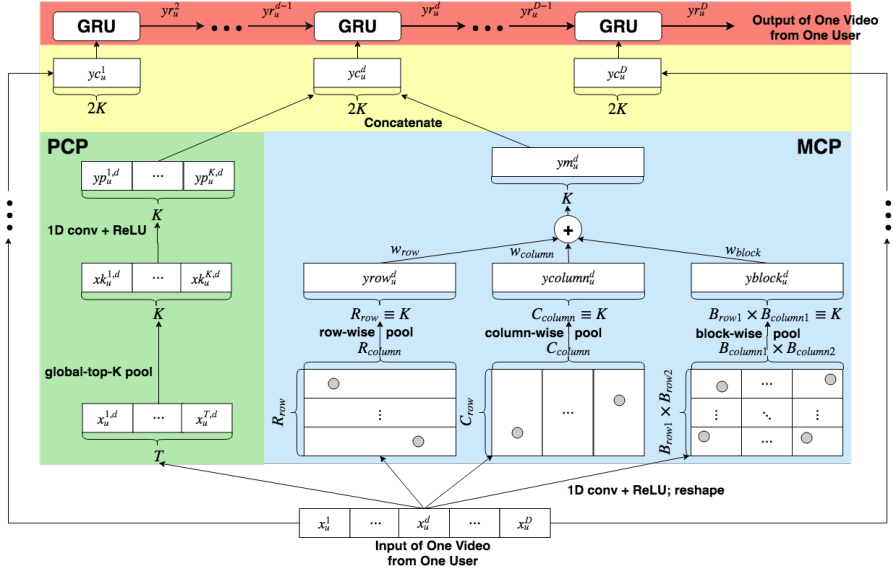}
  \caption{Temporal layers}
\end{figure}

\subsection{Clustering Layers}
The clustering layers process the user dimension of the inputs. There are four components, normalization module, autoencoder module, clustering module and loss module.

The normalization module scales the inputs $T_{(v)}$ according to L2-norm, as $NT_{(v)}$.

The autoencoder module adopts a feedforward structure, which encodes $NT_{(v)}$ to $E_{(v)}$ on the 2D plane $Enc^{2} = (-1.0, 1.0) \times (-1.0, 1.0)$, and then decodes back to $D_{(v)}$. 

In the clustering module, we divide $Enc^{2}$ into $|\Re(C)|$ hierarchical blocks $B^2$. We construct $B^2$ as Figure 4 and follows: 

\begin{enumerate}
\item Let $rintv = (-1.0, 1.0)$, and $NDH$ the number of divisions per right-side hierarchy. Let $RIntv^1=\phi$.
\item Divide $rintv$ uniformly into $NDH$ open sub-intervals $Sub^1$. Let $\alpha= \left(0.0, \frac{1.0}{NDH} \right) \in Sub^1$.  Insert $Sub^1 \setminus \{\alpha\}$ into $RIntv^1$. 
\item Let $rintv=\alpha$. Repeat 2 until 
$|RIntv^1| \geq \frac{1}{2} |\Re(C)|^{1/2}$.
\item Let $Intv^1 = \{ \alpha,(-\alpha_R, -\alpha_L) \rvert $
$\alpha \in RIntv^1 \}$
\item Let $B^2 = Intv^1 \times Intv^1$.
\end{enumerate}

Comparison with K-means results in Figure 4 shows that our supervised-like clustering method maintains the consistency of cluster indices in MBGD training and distributed environment, so that the video cluster can be determined without knowledge of the other ones.

\begin{figure}[ht]
  \centering
  \includegraphics[width=0.47\textwidth]{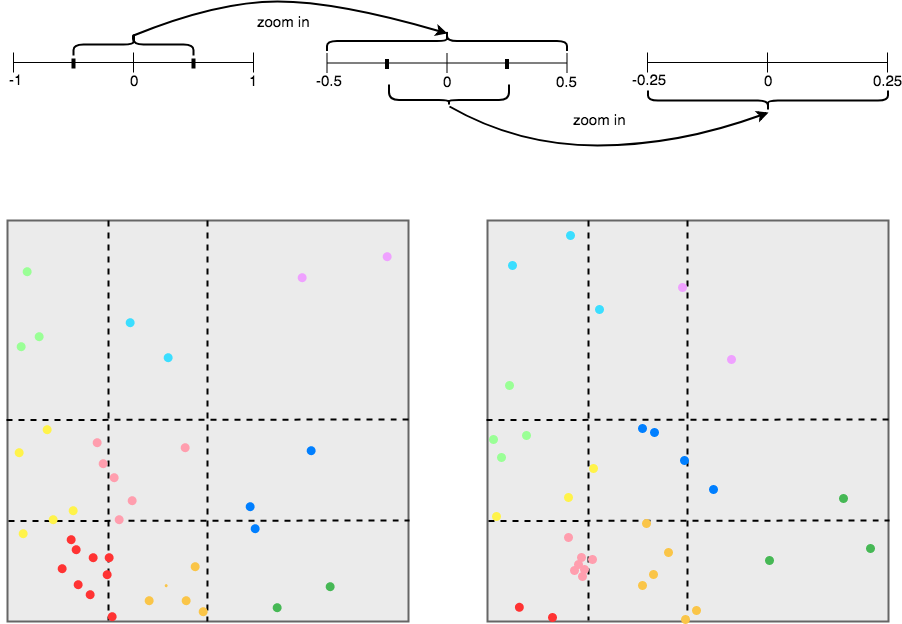}
  \caption{Example of Hierarchical Clustering Blocks}
\end{figure}

The loss module adds the loss from the enc-dec module and the loss from the clustering module weighted by a hyperparameter $\omega$.

$$loss_c = \frac{1}{2} \sum\limits_{v \in V} \left\|NT_{(v)} - D_{(v)} \right\|_2^2 + \omega \min_{b \in B^2} \left\| E_{(v)} - O_b  \right\|_2^2 $$

$O_b$ denotes the coordinate of the center of b.

We apply unsupervised learning to explore the user dimension because there are few labeled tags or accurate prior knowledge. Besides, we separate the temporal layers from the autoencoder module for two reasons: (1) fully-connected structure is not an optimal choice for time series analysis; (2) the decoders with RNN and CNN suffer from lack of a finite-sized dictionary.

\subsection{Accumulation Layers}
The accumulation layers group the primitive inputs by their clusters.

$$XA^c = \sum\limits_{v \in V \bigcap \{C_{(v)} = c\}} X_{(v)}$$

Then follows normalization, and gets $NXA_{(c)}$
After the temporal layers, the outputs are $T_{(c)}$.

\subsection{Policy Layers}
The policy layers estimate the probabilities for each video cluster to be dispatched to all users on day $\emph(D+1)$. There are two components, full connection module and loss module.
The full connection module transforms $T_{(c)}$ to $UP_{(c)}$ on $(0.0, 0.1)^{|U|}$ with learnable parameters.

The loss module compares $UP_{(c)}$ with the real request data on day $\emph{D+1}$.

$$ R_{(c)} = \begin{pmatrix} \frac{XA_1^{D+1}}{\sum\limits_{u \in U}XA_u^{D+1}} & ... & \frac{XA_{|U|}^{D+1}}{\sum\limits_{u \in U}XA_u^{D+1}} \end{pmatrix}_{(c)} $$
$$ loss_p = \frac{1}{2} \sum\limits_{c \in \Re(C)} \left\| UP_{(c)} - R_{(c)} \right\|_2^2 $$

\section{Implementation and Results}

\subsection{Data Set}

We prepare the data set with the real-world video requests from each user accumulated every five minutes. Historical requests are for training, and 15 consecutive days of requests until present are for predicting. There are 177 users, and the dimension of the feature space of one video has 12 minutes $\times$ 24 hours $\times$ 15 days $\times$ 177 users = 764,640.

We analyze the video requests from two aspects. Firstly, we consider each video request sequence from one user as an independent stochastic process and observe the nonstationarities. We sample 5,000 sequences, calculate the difference of the means within a one-day sliding window. After taking the first-order difference of the sequences, the difference of means decreases, as in Table 1. That means nonstationarities are reduced. Next we run the Kwiatkowski-Phillips-Schmidt-Shin (\textbf{KPSS}) test to the sampled sequences with one-day lag (288). They also show the nonstationarities.

\begin{table}[ht]
    \centering
    \begin{tabular}{|c|c|c|}
        \hline
        \multicolumn{3}{|c|}{Difference of Means} \\ 
        \hline
        \multicolumn{2}{|c|}{Original} & 0.00616 \\ \hline
        \multicolumn{2}{|c|}{After first-order differences} &0.00158 \\ \hline
        \multicolumn{3}{|c|}{KPSS Test} \\ \hline
        \multicolumn{2}{|c|}{KPSS statistic} & -5.38 \\ \hline
        \multicolumn{2}{|c|}{P-value} & 0.000003 \\ \hline
        & 1\% & -3.43 \\
        Critical values & 5\% & 2.86 \\
        & 10\% & 2.57\\ \hline
    \end{tabular}
    \caption{Nonstationarity Analysis for Video Requests}
\end{table}

Secondly, we analyze the peak request frequencies. The total request number is 82,700,000 for two hours, with  8,930,000 videos, so the average request frequency per video is 9. We sort the videos in descending order of request frequencies, and show the request distribution after excluding current videoss. Figure 5 shows a typical long-tail shape. 

\begin{figure}[ht]
  \centering
  \includegraphics[width=0.45\textwidth]{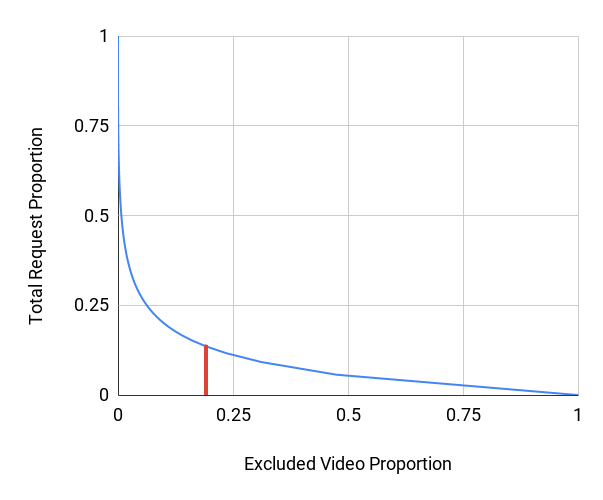}
  \caption{Peak-time Video Request Distribution}
\end{figure}

\subsection{Environment}
We run on multi-machines with 48-core CPU (Intel Xeon 2.20GHz) and 128G memory, based on TensorFlow 1.3 and coordinated with Apache Zookeeper 3.4.

\subsection{Trainers and Predictors}
Training the clustering network requires a clustering trainer, and training the policy network requires a clustering predictor and a policy trainer. Predicting involves clustering and policy predictors.

The trainers comprises all of the modules in the respective networks. The clustering predictor is composed of the all layers but the decoder and loss modules in the clustering network, and the policy predictor is composed of all layers but the loss module in the policy network.

\subsection{Asynchronous Training Mechanism}
Three models are generated through training the temporal, clustering, and policy layers. The clustering model is updated in the clustering network. The temporal and policy models are updated in the policy network. 

We train the two networks asynchronously as in Figure 6. We start the policy trainer first with randomized clustering indices. Then start the clustering predictor and trainer with the identical temporal model trained for several iterations in the policy trainer. The clustering trainer fetches the temporal model from the policy trainer every 200 iterations. The clustering predictor fetches the clustering model from the clustering trainer at each iteration.  

The video dispatch scenario differs from the regular L2R in that the items are assumed as independent in the latter case, while the popular videos often enjoy continual requests. Therefore, the dataset for the policy network needs shuffling to break the chronological order so as to remove autocorrelation between the neighboring samples. This procedure is inspired by experience replay \cite{lin1992,mnih2013} in reinforcement learning. 

All of the learnable parameters are initialized with Xavier initialization \cite{glorot2010}. The MBGD optimizer is utilized, and the learning rates are layer-wisely tuned with decay settings.

\begin{figure}[ht]
  \centering
  \includegraphics[width=0.47\textwidth]{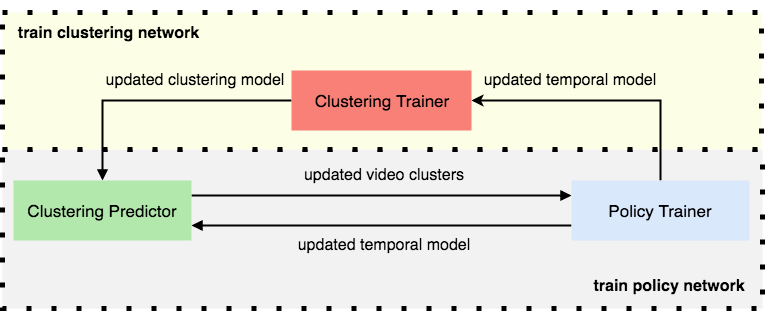}
  \caption{Asynchronous Training Mechanism}
\end{figure}

\subsection{Prediction Mechanism with Post-processing}
\begin{enumerate}
\item Run the clustering predictor for $C(V)$.
\item Run the policy predictor for $UP_{|\Re(C)|,|U|}$.
\item Calculate $CP_{C,I} = UP_{|\Re(C)|,|U|} UC_{|U|,|I|}$.
\item For $\forall i \in I$, sort $V$ in the descending order of $CP_{C(V),i}$. The videos in the same cluster are sorted in the descending order of uploaded time. Dispatch the first 10,000 videos to $i$.
\end{enumerate}

\subsection{Training Performances}
Firstly, Figure 7 shows how the normalization module helps the clustering trainer to converge, where the normalization module is removed for A and B. Loss curve A represents training with a data set of 20 batches, and unchanged temporal models. The loss jumps beyond the initial value immediately after 3 iterations, and then descends slowly. Loss curve B represents training with a data set of only one batch, and periodically (every 500 iteration) changed temporal model. The loss shoots up every time the temporal model is updated to newly randomized values. Loss curve C represents training with a data set of 20 batches, and changed temporal model as in B. It descends 30\% at the first 500 iterations, and continues till the 10,000 iterations.

\begin{figure}[ht]
  \centering
  \includegraphics[width=0.48\textwidth]{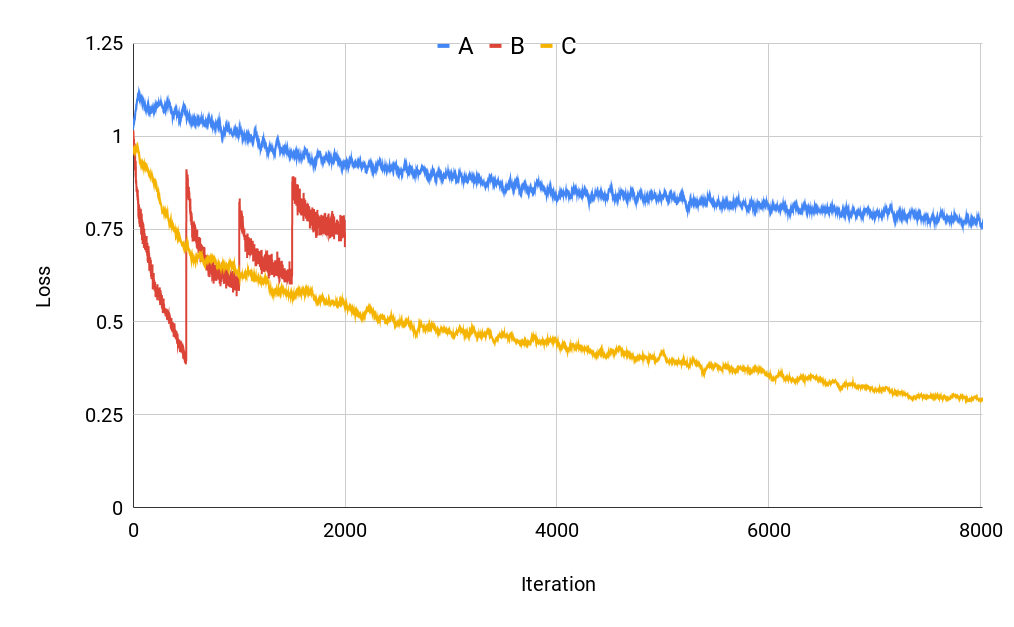}
  \caption{Clustering Trainer Loss Percentages}
\end{figure}

Secondly, Figure 8 shows how shuffled data set helps the policy trainer to converge. The data set in A is ordered by ascending date, and divided into two groups in the middle. The policy trainer fetches the two groups of data alternately, training each group for 200 iterations. The data set in B is finely shuffled. Loss A decreases more rapidly than loss B in the beginning, but shoots up after moving to the other half of data, causing it difficult to converge. 

\begin{figure}[ht]
  \centering
  \includegraphics[width=0.47\textwidth]{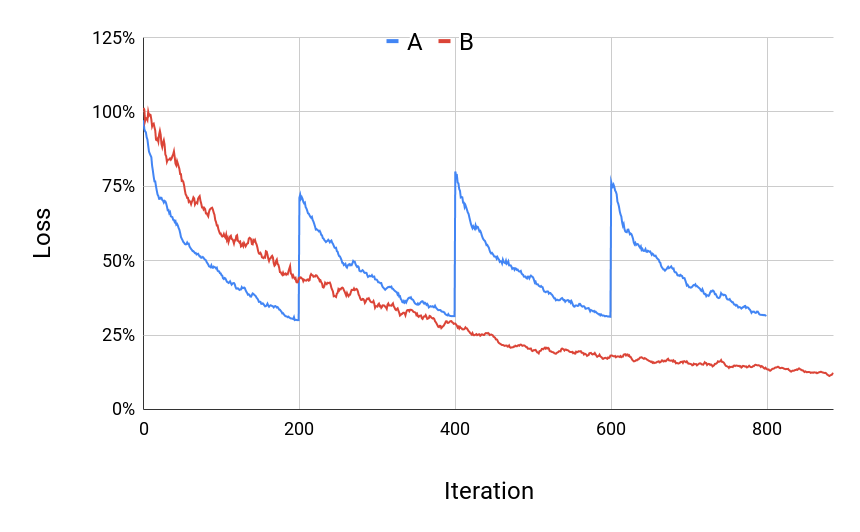}
  \caption{Policy Trainer Loss Percentages}
\end{figure}

Thirdly, Figure 9 shows the training results based on the section of ``asynchronous training mechanism''. They converge without being interfered by each other.

\begin{figure}[ht]
  \centering
  \includegraphics[width=0.46\textwidth]{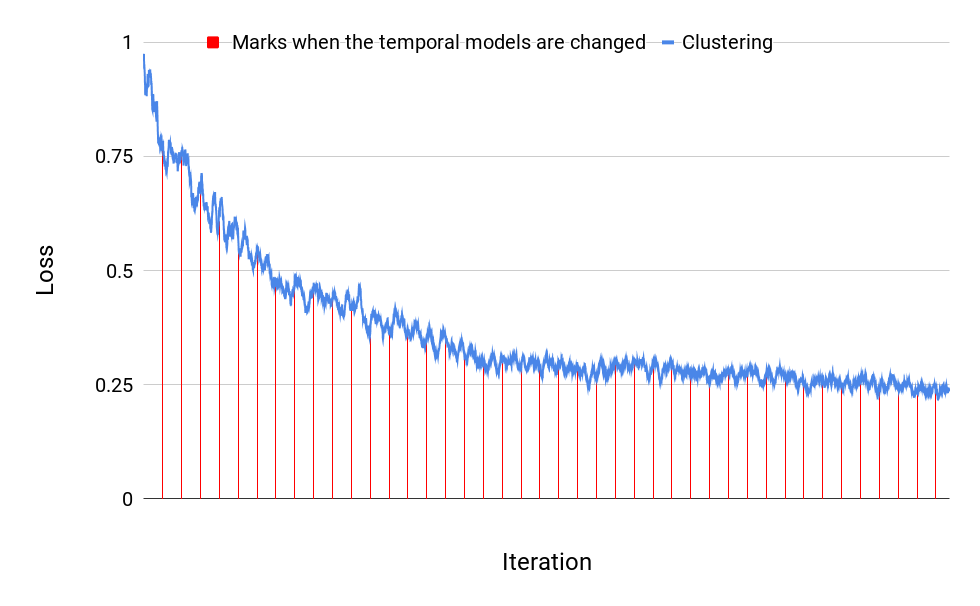}
  \includegraphics[width=0.47\textwidth]{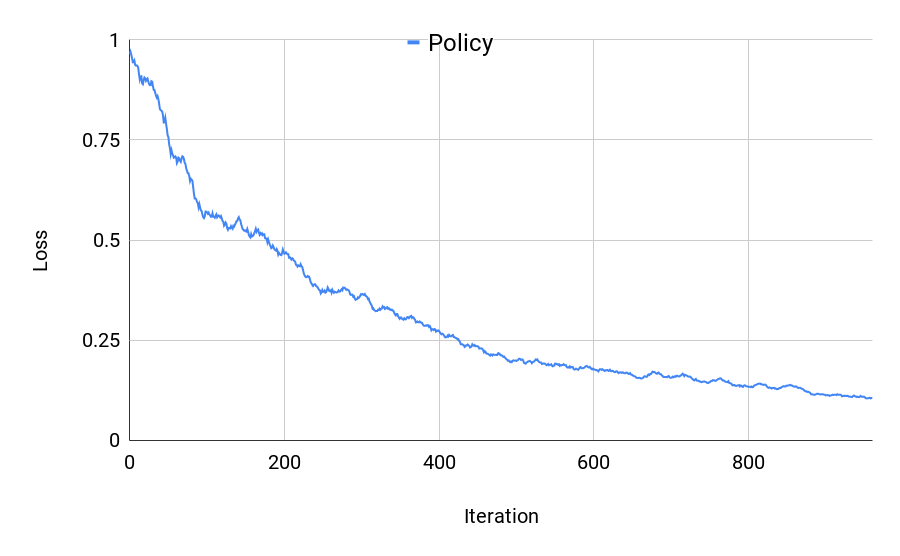}
  \caption{Asynchronous Training Losses of Both Trainers}
\end{figure}

\subsection{Assessment for Video Clustering}
Firstly, Table 2 compares the inner productions between 10,000 video requests in the same clusters (excluding the inner productions of the video requests to themselves) with those in different clusters. The mean from the same clusters is much larger than that from the different, and the coefficient of variation (\textbf{CV}) is smaller. This implies that the video requests from the same cluster are much more similar than from  the different. 

\begin{table}[ht]
    \centering
    \begin{tabular}{|c|c|c|}
        \hline
         & Same & Different \\ \hline
        Mean & 14.0 & 0.117 \\ \hline
        CV & 1.41 & 4.59 \\ \hline
    \end{tabular}
    \caption{Inner Productivities from Same/Different Cluster}
\end{table}

Secondly, Figure 10 compares the video request densities in different clusters, including non-sparsities (\textbf{NS}), L1-norms and L2-norms (whose means are multiplied by 20). NS is defined as the ratio of the number of non-zeros to the length. The means of the three are positively related, and the CVs are close. This implies that the videos with similar request densities are more likely to be in the same cluster.

\begin{figure}[ht]
  \centering
  \includegraphics[width=0.47\textwidth]{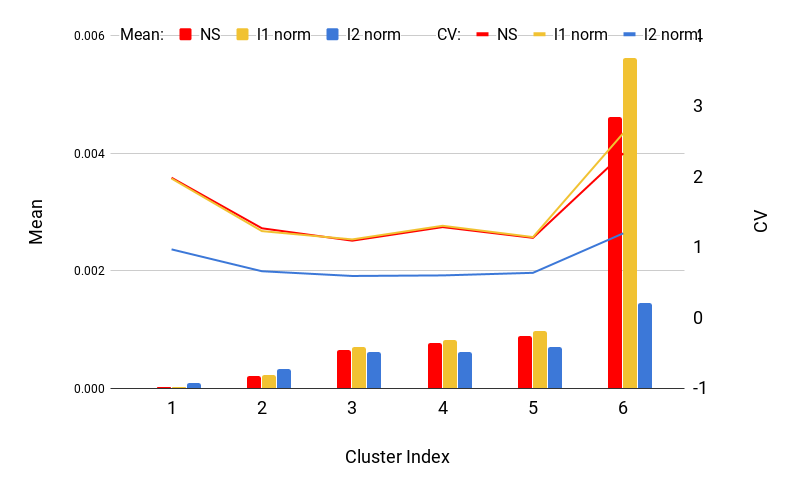}
  \caption{Video Request Densities}
\end{figure}

Thirdly, we count the number of videos (\textbf{NV}) in each cluster, and calculate the average distances (\textbf{AD}) from the encoder outputs to the cluster centroids.Table 3 applies Pearson correlation coefficient to measure the correlation of NV with AD and the area of the cluster. This implies that a larger cluster area for larger values of the encoder outputs is space efficient with sufficient accuracies.

\begin{table}[ht]
    \centering
    \begin{tabular}{|c|c|}
        \hline
        NV and area & 0.6194 \\ \hline
        NV and AD & -0.4956 \\ \hline
    \end{tabular}
    \caption{Correlation coefficient}

\end{table}

\subsection{Prediction and Controlled Experiment Results}
We dispatch respectively 10,000 videos out of 50 million to 50 CDNs at daily off-peak time for the next peak-time requests. Our video dispatch prediction takes 4 hours and 100 machines to compute. We denote our proposed algorithm as A.

We previously dispatched videos based on the threshold of their request frequencies during a period of time, denoted as B.  Since B dispatches videos at peak time, extra costs are not ignorable. The prediction accuracy is represented as the ratio of the peak-time request number (revenues) to the peak-time dispatch number (costs).

For comparison, six other methods are listed. Firstly, we compute dispatch policies via classification \textbf{without video clustering}. This restricts us to build one request model for one user in order to lower the input dimensions. The outputs are either ``to dispatch'' or ``not to dispatch'', while the ground truth is the average requests over the peak time on the next day. Method C applies logistic regression and Method D applies Gradient Boosting Decision Tree.

Secondly, we \textbf{remove the end-to-end learning} mechanism, and let the clustering and policy tasks run independently. This restricts us from using temporal layers in the clustering network. Method E substitutes the clustering network with principal component analysis to reduce the input dimension to 4, followed by K-means clustering. Method F removes the temporal layers from the clustering network.

Thirdly, method G substitutes our supervised-like blocks with \textbf{K-means clustering}. 

Fourthly, method H \textbf{removes the CNN module} from the temporal layers. 

The input sequences in C, D, E, F and H are sampled per hour. The prediction data set in E and G is limited to the latest 10,000 videos, and the dispatch number is 100. The prediction results are shown in Figure 11, which summarize A and B for 90 days, and C to H for 30 days. The results include the means and standard deviations of the prediction accuracies, but exclude recalls because the CDN storage capacities are much smaller than the total size of the requested videos.

The overall performances of the end-to-end learning methods, namely A, G ,H, D and C are better than the others. Although K-means in G and E provides more accurate clustering results, it suffers from scaling issues and limits the dispatch candidates. The CNN module in method A filters the sequences, meanwhile retaining the important information, which especially have advantages over the fully connected structure as in F, and also in D and C. Besides, the CNN module with learned parameters performs better than trivial downsampling method in H. The RNN module in A, G and H provides better sequence forecasting accuracies mainly because RNN is able to explore complex nonlinear state transitions. Lastly, our previous method B suffers from the dilemma of request frequencies and request amounts as in Figure 5. Although a higher threshold in B implies higher dispatch accuracies, the number of dispatchable videos decrease. 

\begin{figure}[ht]
  \centering
  \includegraphics[width=0.5\textwidth]{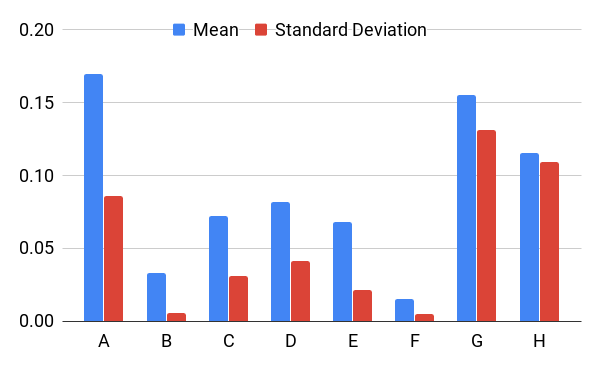}
  \caption{Prediction Results}
\end{figure}

\section{Conclusion}
In this paper, we propose a video dispatch algorithm for VOD to enhance peak-time service quality. We cluster the videos with one network by their request patterns before dispatching them from the large candidate set. We develop dispatch policies for the clustered videos with the other network inspired by pointwise L2R algorithms. The two networks are coupled with shared temporal-feature-extracting layers, which comprise CNN and RNN modules. After training them asynchronously, the average prediction accuracy on real-world video requests is 5 times as high as that of our previous methods.

Future work will be devoted to risk controls to guarantee robustness regardless of rapidly changing situations.

\bibliographystyle{apalike}
\bibliography{arVix}  

\end{document}